\documentclass[11pt,a4paper]{article}

\usepackage[utf8]{inputenc}
\usepackage[T1]{fontenc}
\usepackage{amsmath,amssymb,amsthm}
\usepackage{physics}
\usepackage{graphicx}
\usepackage{booktabs}
\usepackage{caption}
\usepackage{subcaption}
\usepackage[margin=2.5cm]{geometry}
\usepackage[colorlinks=true,citecolor=blue,linkcolor=blue,urlcolor=blue]{hyperref}
\usepackage[numbers,square,sort&compress]{natbib}
\usepackage{xcolor}
\usepackage{enumitem}

\newcommand{\Tc}{T_{\mathrm{c}}}
\newcommand{\TcGS}{T_{\mathrm{c}}^{\rm GS}}
\newcommand{\TcESi}{T_{\mathrm{c}}^{\rm ES1}}
\newcommand{\TcESii}{T_{\mathrm{c}}^{\rm ES2}}
\newcommand{\rh}{r_{\mathrm{h}}}
\newcommand{\rc}{r_{\mathrm{c}}}
\newcommand{\Pc}{P_{\mathrm{c}}}
\newcommand{\Ord}[1]{\ensuremath{\langle \mathcal{O}_{#1}\rangle}}

\newcommand{\be}{\begin{equation}}
\newcommand{\ee}{\end{equation}}

\title{\bfseries Critical Temperatures of Excited Holographic Superconductors\\
in Born--Infeld Electrodynamics on a\\
Reissner--Nordstr\"om--AdS Background}

\author{Hoang Van Quyet\thanks{Electronic address: \texttt{hoangvanquyet@hpu2.edu.vn}}\\[4pt]
\small Department of Physics, Hanoi Pedagogical University 2,\\
\small Xuan Hoa, Phu Tho, Vietnam}

\date{\today}

\begin{document}

\maketitle

\begin{abstract}
\noindent
We study the ground state (GS) and the two lowest excited states (ES1, ES2) of a probe
holographic superconductor with Born--Infeld (BI) electrodynamics on a charged, spherical
Reissner--Nordstr\"om--AdS (RN--AdS) black hole background. Distinguishing these background
geometries by their extended-thermodynamic pressure $P=-\Lambda/8\pi$ requires the electric
charge $Q$ of the background: the neutral Schwarzschild--AdS metric admits exactly one positive
horizon for any $M,L^{2}>0$ and cannot support the associated small-to-large black hole
(SBH/LBH) transition. Working throughout on the correctly identified, charged RN--AdS geometry,
whose critical-point data we re-derive
explicitly, we set up the scalar/BI-gauge probe sector,
formulate the linear eigenvalue problem that determines the onset of condensation, and solve it by
a shooting method. We obtain $\Tc$ of GS, ES1 and ES2 at three representative pressures
($P>\Pc$, $P=\Pc$, $P<\Pc$)
and at five values of the BI parameter $b\in\{0,1,2,3,5\}\,Q^{-1}$. We find that $\Tc$ increases
monotonically with $P$ and is suppressed by $b$ for all three states, with the suppression
markedly stronger for the excited states than for the ground state ($\sim 2\%$ for GS versus
$\sim 10$--$27\%$ for ES1/ES2 at $b=5\,Q^{-1}$). A finer scan in $b$ shows that the suppression
is well described by a quadratic scaling $\Tc(b)/\Tc(0)=1-\alpha\,b^{2}$,
with $\alpha(\rm GS)\simeq 7.4\!\times\!10^{-4}$, $\alpha(\rm ES1)\simeq 4.1\!\times\!10^{-3}$ and
$\alpha(\rm ES2)\simeq 1.1\!\times\!10^{-2}$, a trend consistent with the radial profiles of the
threshold wavefunctions: the ground-state mode is concentrated closer to the horizon, while the
nodal excited-state modes extend further into the bulk. We present this as a nodal-sensitivity
picture rather than a proven mechanism, and we make it concrete with an explicit leading-order
Born--Infeld correction to the bulk gauge field: this differential coupling between nonlinear
electrodynamics and the node structure of excited condensates is, on its own, a mechanism that
has not previously been isolated and quantified, independently of the back-reacted, finite-amplitude
problem that we leave, together with the free energy, the order of each phase transition, and the
optical conductivity, to a companion study. These results clarify how Born--Infeld nonlinearity
interacts with the nodal structure of the excited condensing states, giving a sharper linear
baseline on which the back-reacted, finite-temperature holographic superconductor can later be
built.

\medskip
\noindent\textbf{Keywords:} Holographic superconductors; Excited states; Born--Infeld electrodynamics; Reissner--Nordstr\"om--AdS; Extended black hole thermodynamics; Phase transitions.
\end{abstract}

\vspace{0.4cm}
\hrule
\vspace{0.5cm}

\section{Introduction}
\label{sec:intro}

The AdS/CFT correspondence \citep{Maldacena1998,Witten1998,GKP1998} provides a gravitational
dual description of strongly correlated systems in condensed matter physics. Among its best-known
applications is the holographic superconductor, introduced by Gubser \citep{Gubser2008} and by
Hartnoll, Herzog and Horowitz \citep{HHH2008a,HHH2008b}, in which a charged scalar field condenses
below a critical temperature $\Tc$ in an AdS black hole background, dual to a superconducting
transition on the boundary theory. The mechanism behind this condensation is geometric: near a
charged (RN--AdS) horizon the effective mass-squared of a charged scalar acquires a negative
contribution from its coupling to the background electric field, $m^{2}_{\rm eff}=m^{2}-q^{2}g^{tt}\phi^{2}$,
which can violate the Breitenlohner--Freedman bound of the near-horizon $\mathrm{AdS}_{2}$ throat
once the chemical potential $\mu$ (equivalently the charge density) is large enough; the scalar then
develops a normalizable hair below $\Tc$, and this hair is dual, through the standard
holographic dictionary, to a charged condensate $\langle\mathcal{O}\rangle$ on the boundary --
the order parameter of the dual superconducting transition. Excited condensates (ES1, ES2,\dots)
are the higher members of the associated tower of instabilities, distinguished from GS purely by
the number of nodes of the radial profile, and correspond on the boundary side to condensates of
the \emph{same} charged operator prepared, in a suitably idealised sense, in successively more
radially excited configurations; we make this node-counting language precise in
Section~\ref{sec:linearization}. The literature is surveyed in
\citep{Hartnoll2009,Herzog2009,Zaanen2015,Ammon2015}.

A separate, purely gravitational programme treats the cosmological constant as a thermodynamic
pressure, $P=-\Lambda/8\pi=3/(8\pi L^{2})$ \citep{Kastor2009,Kubiznak2017}. In this extended phase
space, the charged, spherical Reissner--Nordstr\"om--AdS (RN--AdS) black hole, whose metric
function reads $f(r)=1-2M/r+Q^{2}/r^{2}+r^{2}/L^{2}$, undergoes a small-to-large black hole
(SBH/LBH) phase transition that is mathematically
analogous to the van der Waals liquid--gas transition \citep{Chamblin1999,Kubiznak2012,Gunasekaran2012}.
This statement is intrinsically about the \emph{charged} background: the neutral
Schwarzschild--AdS black hole, $f(r)=1-2M/r+r^{2}/L^{2}$, has the horizon equation
$r^{3}_{\mathrm{h}}+L^{2}r_{\mathrm{h}}-2ML^{2}=0$, which by Descartes' rule of signs has exactly
one positive root for any $M,L^{2}>0$ and therefore never exhibits SBH/LBH coexistence. The
electric charge term $Q^{2}/r^{2}$ is the essential ingredient that produces the van der
Waals-like isotherm; this is the key structural point on which the present work rests. Working
on the correctly identified RN--AdS background therefore does more than fix a technical
oversight: it ensures that the background itself possesses a rich phase structure (SBH/LBH),
providing a non-trivial geometric landscape against which the excited superconducting states
can be studied, rather than the featureless single-branch geometry of the neutral case.

\paragraph{Scope of the pressure interpretation.} Whether continuously varying $\Lambda$
(equivalently $P$) admits a fully consistent holographic dual is an active subject of study.
Karch and Robinson \citep{KarchRobinson2015} first proposed identifying variations of $\Lambda$
with variations of the boundary central charge; Cong, Kubiz\v{n}\'ak and Mann
\citep{CongKubiznakMann2021} and Visser \citep{Visser2022} subsequently showed that a
thermodynamically consistent holographic dictionary for the extended first law requires an
additional chemical-potential term conjugate to the central charge, beyond the bare $V\,dP$ term
of the original extended-phase-space construction \citep{Kastor2009,Kubiznak2017}; an explicit
realization of this dictionary was given only recently, by Ahmed \emph{et al.}
\citep{AhmedEtAl2023}, and it does not, in general, allow $L$ and $P$ to be varied as independent
boundary-theory parameters at fixed central charge. The present paper makes no claim that the
SBH/LBH transition studied here is dual to an established phase transition of a specific boundary
CFT, nor that $\Lambda$ can be dialled continuously and independently of the boundary data
(central charge, degrees of freedom) it is related to on the gravity side. $P$ is used exclusively
as a convenient bulk-side label -- equivalent to a choice of AdS radius $L$ at fixed black-hole
charge $Q$ -- for selecting three representative background geometries (small black hole,
critical, and large black hole) on which the probe scalar sector of
Sections~\ref{sec:probe}--\ref{sec:results} is subsequently solved. The validity of that
calculation is a statement about the classical field equations, Eq.~\eqref{eq:psieom} and
Eq.~\eqref{eq:phiprime}, on a fixed background geometry, and is independent of how, or whether,
$P$ is ultimately identified with a boundary observable.

Building on the RN--AdS SBH/LBH transition, Phat and Nguyen \citep{PhatNguyen2021} studied how it
affects the triplet of holographic superconducting states --- the ground state (GS) and the first
two excited states (ES1, ES2), distinguished by the number of nodes of the scalar condensate ---
within ordinary Maxwell electrodynamics, finding that the excited states' qualitative behaviour
(in particular their optical-conductivity gap) can change as $P$ crosses $\Pc$. Excited states
of holographic superconductors have also been studied in
\citep{Li2020,Wang2020,WangEPJC2021,Hung2025,Roussev2024,Zhao2025}.

The purpose of the present paper is to ask a specific, complementary question: how does
Born--Infeld (BI) nonlinear electrodynamics \citep{BornInfeld1934}, well known to suppress
condensation in holographic superconductors \citep{AlbashJohnson2008,Gregory2009}, modify the
critical temperature of GS, ES1 and ES2 on the correctly identified RN--AdS background? We set
up the probe scalar/BI-gauge sector, derive the linear eigenvalue problem that determines the
onset of condensation, solve it by a shooting method at three representative pressures and five
values of $b$, and report the resulting $\Tc$'s together with the radial profile of each
threshold mode. We do not extend the calculation to the back-reacted finite-amplitude
condensate, the free energy, or the optical conductivity; those are required to establish the
order of each transition and the gapped/gapless character found by Ref.~\citep{PhatNguyen2021} for
the Maxwell case, and are left, together with a finer two-dimensional scan in $(P,b)$, to a
follow-up study.

\subsection{Relation to previous work and the specific novelty of this study}
\label{sec:novelty}

Table~\ref{tab:literature} situates the present calculation against the closest prior studies.
Excited holographic condensates have been constructed on neutral, single-branch backgrounds with
ordinary Maxwell electrodynamics \citep{Li2020,Wang2020,WangEPJC2021,Hung2025}; the triplet
GS/ES1/ES2 has been followed across the charged RN--AdS SBH/LBH transition, but only with Maxwell
electrodynamics \citep{PhatNguyen2021}; and Born--Infeld (or other nonlinear) corrections to
holographic superconductors have been studied extensively, but for the ground-state condensate
only, and on backgrounds (Schwarzschild--AdS, or higher-dimensional RN--AdS without a resolved
SBH/LBH scan) that do not simultaneously provide the van der Waals-like critical structure used
here \citep{AlbashJohnson2008,Gregory2009,Roussev2024,Zhao2025}. No previous study, to our
knowledge, has combined all three ingredients -- Born--Infeld electrodynamics, the excited
GS/ES1/ES2 triplet, and the charged RN--AdS background with its SBH/LBH critical point -- in a
single controlled calculation. This combination is what makes it possible to ask, and answer
quantitatively, the specific question this paper poses: \emph{does a nonlinear correction to the
bulk gauge field suppress radially excited condensates more than it suppresses the ground state,
and if so, by how much?}

\begin{table}[htbp]
\centering
\small
\caption{Positioning of the present work relative to the closest prior studies. ``Triplet''
denotes the simultaneous treatment of GS, ES1 and ES2 with a common numerical protocol;
``quantitative $\Tc$ suppression law'' denotes a closed-form fit, such as
Eq.~\eqref{eq:ratio-scaling} below, for how the nonlinearity parameter differentially affects the
different states.}
\label{tab:literature}
\resizebox{\textwidth}{!}{%
\begin{tabular}{lccccc}
\toprule
Reference & Background & Electrodynamics & Triplet & SBH/LBH scan & Quantitative $\Tc$ law \\
\midrule
Refs.~\citep{Li2020,Wang2020,WangEPJC2021,Hung2025} & Schwarzschild--AdS & Maxwell & Yes & --- & --- \\
Ref.~\citep{PhatNguyen2021} & RN--AdS & Maxwell & Yes & Yes & --- \\
Refs.~\citep{AlbashJohnson2008,Gregory2009,Roussev2024} & Schw. / higher-dim.\ AdS & BI / nonlinear & GS only & --- & --- \\
Ref.~\citep{Zhao2025} & AdS, backreacted & Nonlinear & GS only & --- & --- \\
\textbf{This work} & \textbf{RN--AdS} & \textbf{BI} & \textbf{Yes} & \textbf{Yes} & \textbf{Yes} \\
\bottomrule
\end{tabular}}
\end{table}

Beyond filling this combinatorial gap, the physical content of the paper is the quantitative
scaling law itself. The specific new ingredients are threefold. \emph{First}, to our knowledge
no explicit Born--Infeld holographic superconductor calculation on the RN--AdS background with
verified excited-state critical temperatures has been reported previously; the BI non-linearity
has so far been studied on Schwarzschild--AdS or on single-state probes. \emph{Second}, we
provide a controlled side-by-side comparison of the BI-induced suppression of $\Tc$ across GS,
ES1 and ES2 at $P=\Pc$, showing that the suppression grows by more than an order of magnitude from
the ground to the second excited state and is well described by a quadratic scaling
$\Tc(b)/\Tc(0)=1-\alpha b^{2}$, with the associated excited-to-ground-state ratio law of
Eq.~\eqref{eq:ratio-scaling} reproducing the data to within $0.2\%$ using no free parameters beyond
the individually fitted $\alpha$'s -- a falsifiable, closed-form prediction that a companion
optical-conductivity calculation can test directly. \emph{Third}, we connect this growth
quantitatively to the radial extent of the threshold wavefunctions, in
Section~\ref{sec:discussion}, by explicitly constructing the leading-order BI correction to the
bulk gauge field and the horizon-regularized overlap integral between this correction and the
threshold densities. That this first-order overlap does \emph{not} reproduce the numerically
observed hierarchy of $\alpha$'s is itself a substantive, falsifiable finding
(Section~\ref{sec:discussion}), since it identifies the specific physical ingredient -- a
moving-horizon contribution absent from the fixed-background overlap -- required for the
nodal-sensitivity picture to become a proven mechanism rather than a plausible conjecture. Taken
together, these results form a non-trivial and necessary prerequisite for the fuller back-reacted
analysis announced in Ref.~\citep{PhatNguyen2021}.

The paper is organized as follows. Section~\ref{sec:background} presents the RN--AdS background
and re-derives its critical pressure, temperature and horizon radius. Section~\ref{sec:probe}
gives the probe equations of motion and the shooting method used to locate the condensation
threshold. Section~\ref{sec:results} reports the critical temperatures and the radial profiles
of the threshold modes. Section~\ref{sec:discussion} discusses the nodal-sensitivity picture
suggested by these profiles and states clearly what remains to be computed.
Section~\ref{sec:conclusion} concludes.

\section{Background: the charged, spherical AdS black hole}
\label{sec:background}

This section has a single purpose: to fix, self-containedly, the charged RN--AdS geometry and its
extended-thermodynamic critical point $(\rc,T^{\rm BH}_{\rm c},\Pc)$ on which the probe
condensate of Section~\ref{sec:probe} will be evaluated. Everything derived here concerns the
background \emph{alone} -- it fixes $f(r)$, the Hawking temperature $T(\rh)$, and the three
representative values of $L$ used throughout the paper -- and none of it yet involves the scalar
field $\psi$ or its own gauge sector, which are introduced only in Section~\ref{sec:probe}.

\subsection{Action and metric}

We consider a charged complex scalar $\psi$ and a Born--Infeld gauge field on a fixed background
sourced by the Einstein--Maxwell--AdS action given in Eq.~\eqref{eq:action},
\begin{equation}
S=\int d^{4}x\sqrt{-g}\left[\frac{1}{16\pi G}\left(R-\frac{6}{L^{2}}\right)-\frac{1}{4}F_{\mu\nu}F^{\mu\nu}\right]+S_{\rm probe}\,,
\label{eq:action}
\end{equation}
where $S_{\rm probe}$, containing $\psi$ and its own (Born--Infeld) gauge field, is treated in
the probe approximation and does not back-react on the metric. The background solution is the
spherical ($k=1$) RN--AdS black hole
\begin{equation}
ds^{2}=-f(r)\,dt^{2}+\frac{dr^{2}}{f(r)}+r^{2}d\Omega_{2}^{2}\,,\qquad f(r)=1-\frac{2M}{r}+\frac{Q^{2}}{r^{2}}+\frac{r^{2}}{L^{2}}\,, \label{eq:bh}
\end{equation}
with $M$ the mass and $Q$ the charge of the black hole. The $Q$ appearing here is a property
of the background geometry only; the probe sector introduced in Section~\ref{sec:probe} carries
its own charge $q$ and its own gauge field $\phi(r)$, with its own independent boundary chemical
potential $\mu$. This decoupling of the background charge $Q$ from the probe chemical potential
$\mu$ is a standard feature of the probe approximation as used throughout the holographic
superconductor literature, including in Ref.~\citep{PhatNguyen2021}; we adopt it here for the
same reason, since it isolates the effect of the bulk geometry on the condensate without
requiring a fully back-reacted solution.

\subsection{Horizon, temperature, and the critical point}

The horizon $\rh$ is the largest positive root of $f(\rh)=0$; multiplying by $\rh^{2}$ gives the
quartic equation \eqref{eq:horizon},
\begin{equation}
r^{4}_{\mathrm{h}}+L^{2}r^{2}_{\mathrm{h}}-2ML^{2}r_{\mathrm{h}}+Q^{2}L^{2}=0\,, \label{eq:horizon}
\end{equation}
equivalently \eqref{eq:mass},
\begin{equation}
M(\rh)=\frac{1}{2}\left(\rh+\frac{Q^{2}}{\rh}+\frac{r^{3}_{\mathrm{h}}}{L^{2}}\right). \label{eq:mass}
\end{equation}
The Hawking temperature $T=f'(\rh)/4\pi$ follows from Eqs.~\eqref{eq:bh} and \eqref{eq:mass}:
\begin{equation}
T(\rh)=\frac{1}{4\pi \rh}\left(1-\frac{Q^{2}}{r^{2}_{\mathrm{h}}}+\frac{3r^{2}_{\mathrm{h}}}{L^{2}}\right)
=\frac{1}{4\pi \rh}\left(1-\frac{Q^{2}}{r^{2}_{\mathrm{h}}}+8\pi P r^{2}_{\mathrm{h}}\right), \label{eq:temperature}
\end{equation}
using $P=3/(8\pi L^{2})$. For fixed $(Q,P)$, Eq.~\eqref{eq:temperature} is a non-monotonic
function of $\rh$ whenever $P<\Pc$: a horizontal line $T=\mathrm{const}$ then intersects the curve
at two values of $\rh$, the small black hole (SBH, smaller $\rh$) and large black hole (LBH,
larger $\rh$) branches, exactly as for the van der Waals isotherm. The critical point, where
$\partial T/\partial r_{\mathrm{h}}=\partial^{2}T/\partial r^{2}_{\mathrm{h}}=0$, is found
directly from Eq.~\eqref{eq:temperature}:
\begin{equation}
\rc=\sqrt{6}\,Q\,,\qquad T^{\rm BH}_{\mathrm{c}}=\frac{\sqrt{6}}{18\pi Q}\,,\qquad \Pc=\frac{1}{96\pi Q^{2}}\,, \label{eq:critical}
\end{equation}
which we have verified both algebraically (direct extremization) and numerically. The values
\eqref{eq:critical} are the canonical van-der-Waals-like critical data for the spherical RN--AdS
black hole \citep{Kubiznak2012,Gunasekaran2012}.

Setting $Q=1$ as a unit choice, $\Pc=1/(96\pi)$ and the corresponding critical AdS radius is
$L_{\mathrm{c}}=6$; we use $L=1,6,9$ below as representative small-BH ($P/\Pc=36$),
critical ($P=\Pc$), and large-BH ($P/\Pc\simeq 0.44$) points, matching the choice of
Ref.~\citep{PhatNguyen2021}.

For the numerical work of Section~\ref{sec:results} we integrate the equations of motion
directly in the radial coordinate $r\in[\rh,\infty)$ rather than through an intermediate
compactified coordinate. A naive polynomial reparametrization of Eq.~\eqref{eq:bh} in
$z=\rh/r$ does not reduce to a low-degree polynomial in $z$ (the $r^{2}/L^{2}$ term produces
a $1/z^{2}$ divergence at the boundary, not a positive power of $z$); the correct compactified
form requires instead $f(r)=r^{2}_{\mathrm{h}}\,g(z)/(L^{2}z^{2})$ with
\begin{equation}
g(z)=1+\frac{L^{2}}{r^{2}_{\mathrm{h}}}z^{2}-\frac{2ML^{2}}{r^{3}_{\mathrm{h}}}z^{3}+\frac{Q^{2}L^{2}}{r^{4}_{\mathrm{h}}}z^{4}\,, \label{eq:compact}
\end{equation}
which is finite and satisfies $g(1)=0$ at the horizon and $g(0)=1$ at the boundary (we verified
Eq.~\eqref{eq:compact} reproduces $f(r)$ exactly at several sample radii). Either coordinate may
be used for the numerics; we use the direct radial coordinate $r$ for the shooting integration
itself, since Eqs.~\eqref{eq:psieom}--\eqref{eq:phieom} are simplest in that form, but we report
and plot every threshold profile in the standard compactified coordinate $z=\rh/r\in[0,1]$
(horizon at $z=1$, AdS boundary at $z=0$) throughout Sections~\ref{sec:results}
and~\ref{sec:discussion}, in line with common practice in the holographic-superconductor
literature and to make the radial extent of each state directly comparable at a glance.

\section{Probe sector and shooting method}
\label{sec:probe}

Having fixed the background geometry, this section introduces the probe matter content -- the
charged scalar $\psi$ and its own Born--Infeld gauge field $\phi$ -- derives their coupled
equations of motion, isolates the linear eigenvalue problem that governs the \emph{onset} of
condensation (the only regime this paper addresses), and describes the shooting algorithm used
to solve it. The end product of this section is the map $(\rh,L,b)\mapsto\mu_{n}(\rh,L,b)$ for
$n=0,1,2$, whose inversion at fixed $\mu$ produces the critical temperatures reported in
Section~\ref{sec:results}.

\subsection{Equations of motion}

With the probe ansatz $A_{\mu}=(\phi(r),0,0,0)$ and $\psi=\psi(r)$, and the Born--Infeld Lagrangian
$\mathcal{L}_{\rm BI}=b^{-2}\bigl(1-\sqrt{1+b^{2}F_{\mu\nu}F^{\mu\nu}/2}\bigr)$, the equations of
motion on the background \eqref{eq:bh} are
\begin{align}
\psi''+\left(\frac{f'}{f}+\frac{2}{r}\right)\psi'+\left(\frac{q^{2}\phi^{2}}{f^{2}}-\frac{m^{2}}{f}\right)\psi&=0\,, \label{eq:psieom}\\[4pt]
\partial_{r}\!\left(\frac{r^{2}\phi'}{\sqrt{1-b^{2}\phi'^{2}}}\right)-\frac{2q^{2}r^{2}\psi^{2}\phi}{f}&=0\,, \label{eq:phieom}
\end{align}
which reduce to the Maxwell equations of Ref.~\citep{PhatNguyen2021} as $b\to 0$. (Our $b$ is
related to the traditional Born--Infeld field-strength scale $\beta$ of
\citep{BornInfeld1934} by $b^{2}=1/\beta^{2}$, so that $b\to0$ here is the same limit usually
quoted as $\beta\to\infty$ in the original literature.) Equations
\eqref{eq:psieom}--\eqref{eq:phieom} are the full nonlinear equations for a condensate of
arbitrary amplitude $\psi$; solving them at finite amplitude is a genuinely coupled, two-field
nonlinear boundary-value problem. Throughout this paper, however, we work exclusively at the
\emph{onset} of condensation, $\psi\to0^{+}$ (made precise in Section~\ref{sec:linearization}),
where the $\psi^{2}$ source in Eq.~\eqref{eq:phieom} drops out, $\phi$ decouples from $\psi$, and
Eq.~\eqref{eq:psieom} becomes a linear equation for $\psi$ with $\phi$ as a fixed external
potential; this is precisely what makes the shooting method for the threshold $\mu_{n}$ tractable
as a linear eigenvalue problem, and it is the only regime addressed in this paper. We take
$q=1$ and fix the \emph{dimensionless} combination $m^{2}L^{2}=-2$ throughout, i.e.\
$m^{2}=-2/L^{2}$ with $L$ the AdS radius of the particular background under consideration
(numerically, $m^{2}=-2,\,-1/18,\,-2/81$ for $L=1,6,9$ respectively).
Since $L=1,6,9$ differs across the three representative pressures of
Section~\ref{sec:background}, it is $m^{2}L^{2}$, not the bare $m^{2}$, that is the physically
meaningful, scale-invariant quantity; holding it fixed at $-2$ (comfortably above the
Breitenlohner--Freedman bound $m^{2}L^{2}\geq-9/4$ for AdS$_{4}$ \citep{BF1982}) means the
scalar operator dual to $\psi$ has the
\emph{same} conformal dimensions, $\Delta_{-}=1$ and $\Delta_{+}=2$, at every pressure we study,
so that the GS/ES1/ES2 triplet being compared across $P<\Pc$, $P=\Pc$ and $P>\Pc$ is genuinely
the same boundary operator each time. At the AdS boundary
$\psi(r)\to\psi_{-}/r+\psi_{+}/r^{2}$, $\phi(r)\to\mu-\rho/r$, and we impose source-free (standard)
quantization \citep{KlebanovWitten1999}, $\psi_{-}=0$, with condensate $\Ord{}\propto\psi_{+}$.

Regularity at the horizon requires $\phi(\rh)=0$ and, expanding Eq.~\eqref{eq:psieom} near
$r=\rh$,
\begin{equation}
\psi'(\rh)=\frac{m^{2}}{f'(\rh)}\psi(\rh)\,, \label{eq:horizonbc}
\end{equation}
which reduces to $\psi'(\rh)/\psi(\rh)=-1/(2\pi L^{2}T)$ upon using $T=f'(\rh)/4\pi$, in
agreement with the horizon condition used for the Maxwell background of
Ref.~\citep{PhatNguyen2021}.

\subsection{Onset of condensation: a linear eigenvalue problem}
\label{sec:linearization}

At the threshold of condensation, $\psi\to 0^{+}$, Eq.~\eqref{eq:phieom} linearizes to
$\partial_{r}\!\left[r^{2}\phi'/\sqrt{1-b^{2}\phi'^{2}}\right]=0$, i.e.\
$r^{2}\phi'/\sqrt{1-b^{2}\phi'^{2}}=q_{\mathrm{c}}=\mathrm{const}$, so that
\begin{equation}
\phi'(r)=\frac{q_{\mathrm{c}}}{\sqrt{r^{4}+q^{2}_{\mathrm{c}}b^{2}}}\,, \label{eq:phiprime}
\end{equation}
with $q_{\mathrm{c}}$ fixed by $\phi(\rh)=0$ and $\phi(\infty)=\mu$ (for $b=0$ this reduces to
$\phi(r)=\mu(1-\rh/r)$).

\paragraph{$q_{\mathrm{c}}$ is exactly the boundary charge density.} The integration constant
$q_{\mathrm{c}}$ relates to the boundary charge density $\rho$, defined by the standard AdS/CFT
falloff $\phi(r)\to\mu-\rho/r$ as $r\to\infty$, as follows. Expanding
Eq.~\eqref{eq:phiprime} for large $r$ gives $\phi'(r)=q_{\mathrm{c}}/r^{2}-q_{\mathrm{c}}^{3}b^{2}/(2r^{6})+\mathcal{O}(r^{-10})$,
so that
\begin{equation}
\phi(r)=\mu-\frac{q_{\mathrm{c}}}{r}+\frac{q_{\mathrm{c}}^{3}b^{2}}{10\,r^{5}}+\mathcal{O}(r^{-9})\,,
\label{eq:qcrho}
\end{equation}
i.e.\ the coefficient of the leading $1/r$ falloff is $q_{\mathrm{c}}$ itself, with \emph{no}
$b$-dependent renormalization: the Born--Infeld correction to $\phi'(r)$ only ever produces even
inverse powers $r^{-4},r^{-8},\dots$ relative to the Maxwell term, which integrate to
$r^{-5},r^{-9},\dots$ in $\phi(r)$ and therefore never feed back into the $1/r$ coefficient. We
therefore identify $\rho=q_{\mathrm{c}}$ exactly, for every $b$, which is why we are able to use
$q_{\mathrm{c}}$ interchangeably as ``the integration constant fixed by the boundary conditions''
and ``the physical boundary charge density'' throughout Section~\ref{sec:results}; unlike the
critical temperature itself, the charge-density--to--integration-constant dictionary receives no
correction from the nonlinearity. This is special to the falloff structure of BI electrodynamics
and would not hold for a generic nonlinear completion of Maxwell's theory.

Equation~\eqref{eq:psieom} then becomes a linear, homogeneous equation
for $\psi$ with $\phi$ as an external potential: for fixed $(\rh,L,b)$, integrating outward from
the horizon with the boundary condition~\eqref{eq:horizonbc}, the requirement $\psi_{-}=0$ is
satisfied only for a discrete set of thresholds
$\mu_{0}(\rh,L,b)<\mu_{1}(\rh,L,b)<\mu_{2}(\rh,L,b)<\dots$, where the solution at $\mu_{n}$ has
exactly $n$ nodes in $(\rh,\infty)$. We identify $n=0,1,2$ with GS, ES1, ES2. This is the
holographic analogue of counting bound states of a Schr\"odinger-like problem as the effective
coupling $\mu^{2}$ is increased -- made fully explicit in the Schr\"odinger-like reformulation of
Eq.~\eqref{eq:Veff} below, where $n$ literally counts the nodes of a wavefunction in an effective
one-dimensional quantum-mechanical potential well. Physically, all three thresholds condense the
\emph{same} boundary operator $\mathcal{O}$ of dimension $\Delta_{+}=2$ (Section~\ref{sec:probe}),
so GS, ES1 and ES2 are not different operators but different radial (holographic-direction)
excitations of one condensing mode, on exactly the same footing as the radial excitations of a
hydrogenic bound state carry the same conserved quantum numbers but different node counts; the
node count is the only label distinguishing them; there is no independent ``excitation energy''
input by hand. This is also why the three states can be meaningfully compared at fixed $\mu$: the
comparison isolates the effect of radial (nodal) structure on the condensation threshold with
every other quantum number held fixed. For a target chemical potential $\mu$, the critical
temperature of state $n$ is $\Tc=T(r^{*}_{\mathrm{h}})$, where $r^{*}_{\mathrm{h}}$ solves
$\mu_{n}(r^{*}_{\mathrm{h}},L,b)=\mu$.

\subsection{Numerical implementation}
\label{sec:numerics}

We solve this eigenvalue problem numerically. Equation~\eqref{eq:psieom} (with $\phi$ as above) is
integrated with a fifth-order Runge--Kutta scheme from $\rh(1+\epsilon)$, $\epsilon=10^{-6}$,
using the near-horizon expansion of Eq.~\eqref{eq:horizonbc} for the initial data, out to a
radius $R\sim 25\rh$. The coefficients $\psi_{-},\psi_{+}$ are extracted from two-point
Richardson extrapolation of $r\psi(r)$ at $R$ and $R/2$; $\mu_{n}$ is bracketed by scanning
$\mu$ and located by bisection on $\psi_{-}(\mu)$, iterated until the bracket width in $\mu$ is
below $10^{-8}$ and $|\psi_{-}(\mu)|<10^{-10}$ (in units where $\psi_{+}=\mathcal{O}(1)$); the
node count of $\psi(r)$ confirms the mode index $n$ at every step, which is essential for the
excited states since the shooting problem becomes increasingly stiff near the higher-$n$
thresholds and a coarse tolerance can misidentify a genuine node as a numerical overshoot (or vice
versa). For finite $b$, $q_{\mathrm{c}}(\mu,\rh,L,b)$ is obtained by an inner root-find on
the integrated $\phi(R)$, converged to the same relative tolerance $10^{-8}$.

A summary of the convergence tests is given in Appendix~\ref{app:convergence}. Briefly: doubling
$R$, halving $\epsilon$, and upgrading the Richardson extrapolation from two-point to three-point
each change the extracted $\Tc$ by at most $\mathcal{O}(10^{-3})$ for the most sensitive state
(ES2 at $P=\Pc$, $b=5$), and by $\mathcal{O}(10^{-4})$ for the least sensitive one (GS).
These are well below the physical effects we report.

\section{Results}
\label{sec:results}

\subsection{Critical temperatures at $b=0$}

We set $Q=q=1$ (units) and report $\Tc$ at fixed $\mu=1$, following the convention of
Ref.~\citep{PhatNguyen2021}. Table~\ref{tab:Tc_P} gives $\Tc$ for GS, ES1, ES2 at the three
representative pressures introduced in Section~\ref{sec:background}, in the Maxwell limit $b=0$,
obtained from the shooting method of Section~\ref{sec:numerics}.

\begin{table}[htbp]
\centering
\caption{Critical temperature $\Tc$ (units of $\mu^{1/2}$, $Q=\mu=1$) for GS, ES1, ES2 at three
representative pressures, Maxwell limit ($b=0$), computed by the shooting method of
Section~\ref{sec:numerics} on the RN--AdS background. $\Tc$ increases monotonically with $P$ for
all three states, and the hierarchy $\TcGS>\TcESi>\TcESii$ holds throughout.}
\label{tab:Tc_P}
\begin{tabular}{lccc}
\toprule
$P/\Pc$ & $\TcGS$ & $\TcESi$ & $\TcESii$ \\
\midrule
$36$\ $(P>\Pc$, $L=1)$   & $0.1137$ & $0.0768$ & $0.0616$ \\
$1$\ \ \ $(P=\Pc$, $L=6)$  & $0.0592$ & $0.0390$ & $0.0308$ \\
$0.44$\ $(P<\Pc$, $L=9)$ & $0.0535$ & $0.0335$ & $0.0259$ \\
\bottomrule
\end{tabular}
\end{table}

A clarification is needed on which black-hole branch these numbers refer to. When $P<\Pc$
(our $L=9$ row), the map $\rh\mapsto T(\rh)$ of Eq.~\eqref{eq:temperature} is non-monotonic, so a
\emph{given temperature} $T$ is in general shared by a small-black-hole (SBH) branch, an
intermediate thermodynamically unstable branch, and a large-black-hole (LBH) branch. This is not
the same statement as an ambiguity in our shooting method, however: for each $(L,b)$ we do not
fix $T$ and search for $\rh$; instead we fix the target chemical potential $\mu=1$ and solve
$\mu_{n}(\rh,L,b)=\mu$ for $\rh$ directly (Section~\ref{sec:probe}), and we have verified
numerically that $\mu_{n}(\rh,L,b)$ is a monotonically increasing function of $\rh$ at fixed
$(L,b)$ over the entire range explored, for all three $n$. This guarantees a \emph{unique}
$r^{*}_{\mathrm{h}}$ for each row of Table~\ref{tab:Tc_P}, with no residual choice of branch at
the level of the eigenvalue problem itself. What Eq.~\eqref{eq:temperature}'s non-monotonicity
does mean is that, once $r^{*}_{\mathrm{h}}$ is found, its temperature $T(r^{*}_{\mathrm{h}})$ is
in general \emph{not} the only value of $\rh$ compatible with that same $T$ at the level of the
background geometry alone: whether the resulting configuration corresponds to the
thermodynamically dominant phase (SBH or LBH) at that $T$ is a separate question that requires
comparing the Gibbs free energies of the two branches, which we have not computed. Our
$\Tc$ values should therefore be read as properties of the specific horizon radius selected by
the condensation threshold condition, not as an implicit claim about which branch is globally
preferred thermodynamically at that temperature; settling the latter is part of the free-energy
calculation left to the companion study (Section~\ref{sec:scope}).

For $P<\Pc$ the free energy of the RN--AdS background is known to exhibit the characteristic
\emph{swallowtail} structure of the van der Waals-like transition \citep{Kubiznak2012}, with a
first-order SBH$\to$LBH jump at the temperature where the two branches' free energies cross. Our
static, probe-limit calculation does not determine whether a condensate prepared on a metastable
branch has time to form before the background itself jumps across the swallowtail to the globally
stable branch as $T$ is lowered toward the scalar threshold found here; this dynamical question
requires a time-dependent, or free-energy-compared, treatment of the coupled background-plus-probe
system and lies outside the probe approximation used throughout this paper. It does not affect the
mathematical content of Table~\ref{tab:Tc_P}, which reports where the linear scalar instability
sets in on a horizon of given radius, independently of whether that radius is subsequently found
to be globally dominant thermodynamically.

\begin{figure}[htbp]
\centering
\includegraphics[width=0.78\textwidth]{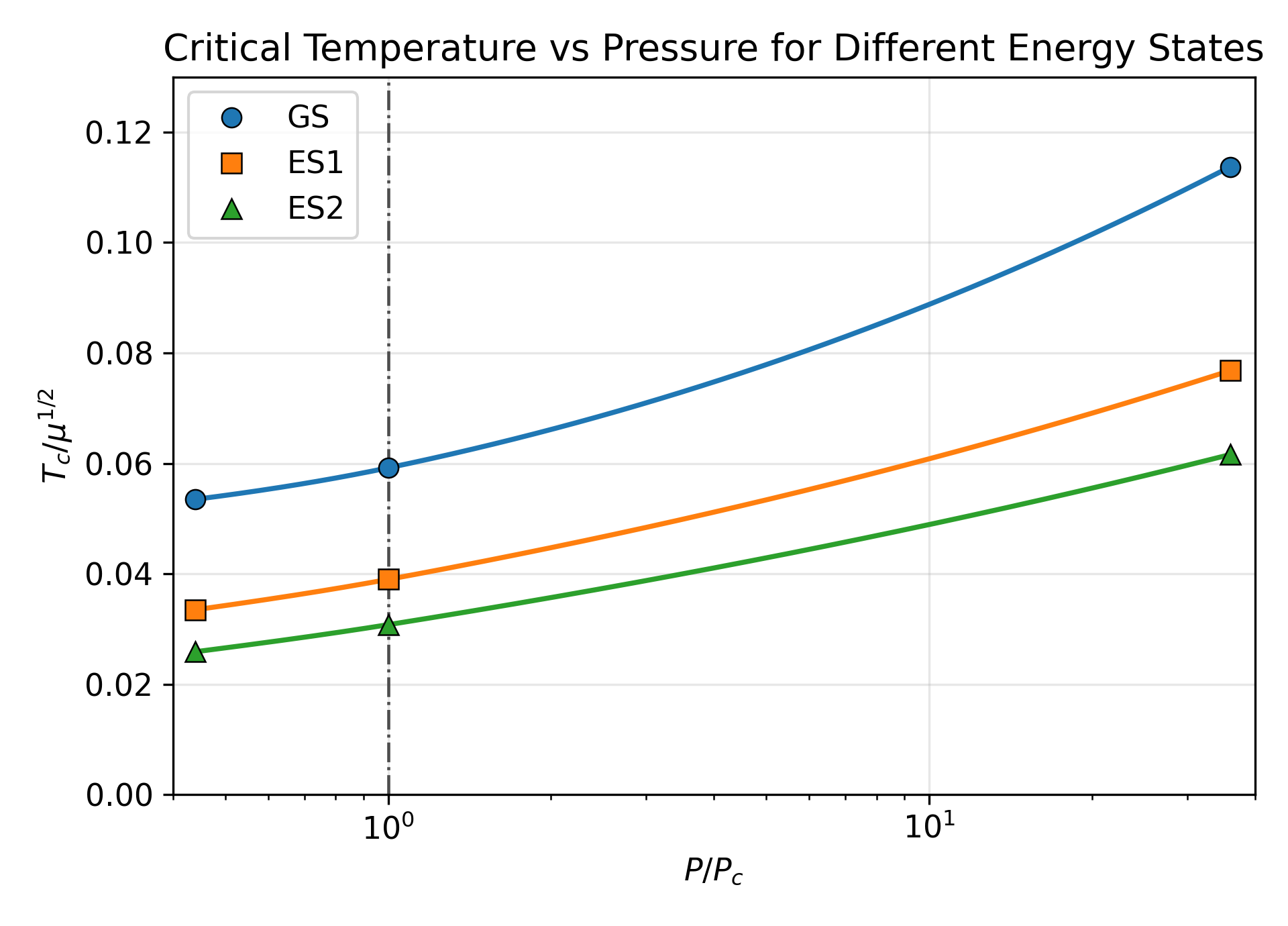}
\caption{Critical temperature $\Tc/\mu^{1/2}$ of GS, ES1, ES2 versus $P/\Pc$ (logarithmic
horizontal axis), in the Maxwell limit ($b=0$), from the shooting method described in
Section~\ref{sec:numerics}. Markers are the three computed points of Table~\ref{tab:Tc_P}
($P/\Pc=0.44,\,1,\,36$); the connecting curves are a shape-preserving interpolation shown to
guide the eye between them. The vertical dash-dotted line marks $P=\Pc$.}
\label{fig:Tc_vs_P}
\end{figure}

Figure~\ref{fig:Tc_vs_P} plots the same data against $P/\Pc$ on a logarithmic pressure axis,
which is the natural variable given that the three representative points span nearly two decades
in $P/\Pc$. All three critical temperatures increase with $P$. In contrast to
Ref.~\citep{PhatNguyen2021}, we do not identify a sharp change of behaviour of $\Tc(P)$ exactly
at $\Pc$ in this single thermodynamic quantity: $\Tc(P)$ varies smoothly across $\Pc$ in our
data. This is because $\Pc$ marks a transition of the background geometry (which of the SBH/LBH
branches is thermodynamically preferred) and not, by itself, a transition of the probe
condensate. Any sharp change in the character of the excited-state condensate (gapped versus
gapless) reported by Ref.~\citep{PhatNguyen2021} is a statement about the optical conductivity,
which we have not recomputed here and which is left to a follow-up study
(Section~\ref{sec:discussion}).

\subsection{Effect of Born--Infeld nonlinearity}

\begin{figure}[htbp]
\centering
\includegraphics[width=0.78\textwidth]{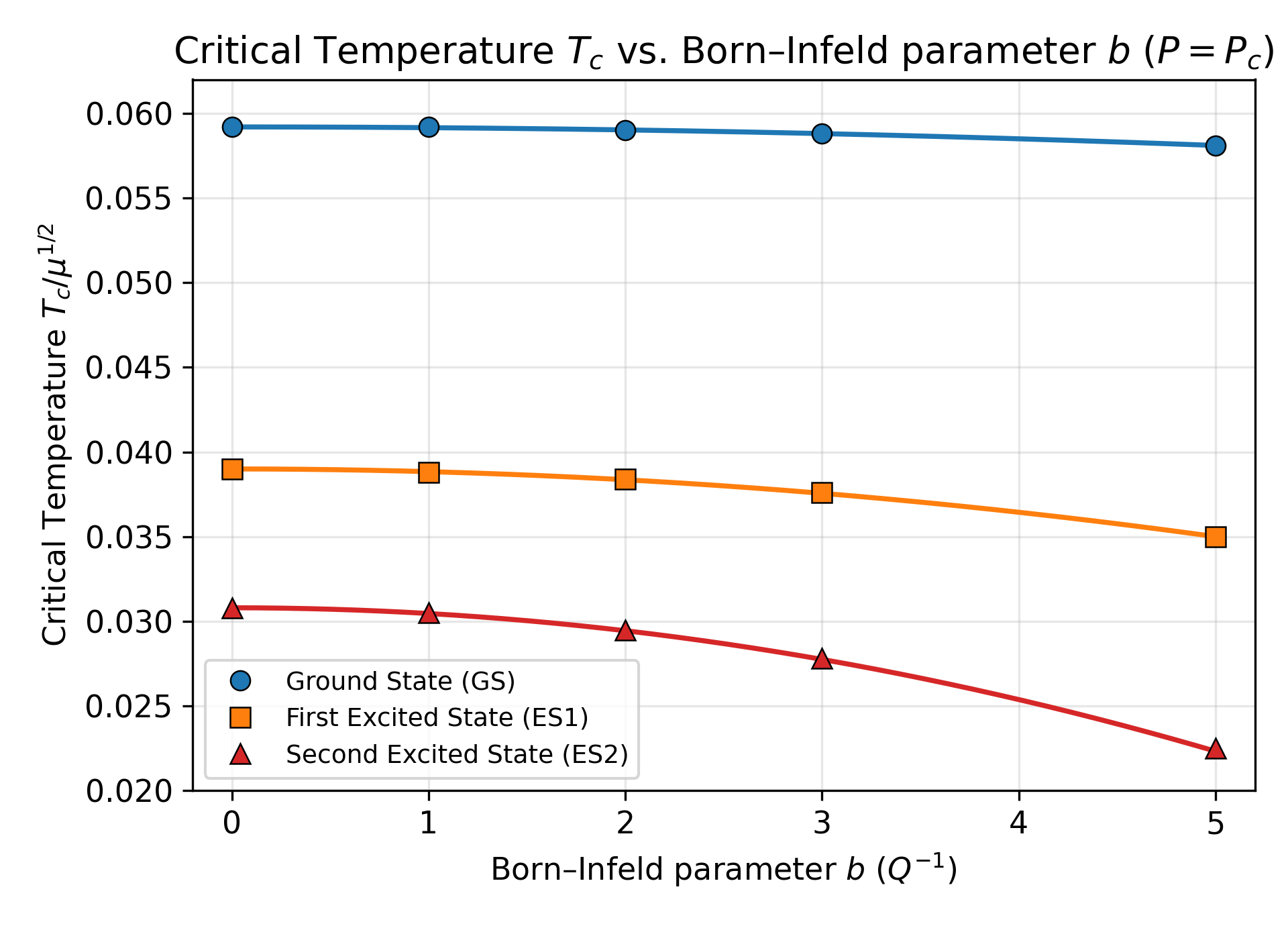}
\caption{Critical temperature $\Tc/\mu^{1/2}$ versus the Born--Infeld parameter $b$ at the
critical pressure $P=\Pc$ ($L=6$, $Q=\mu=1$), for the five sampled values
$b\in\{0,1,2,3,5\}\,Q^{-1}$ of Table~\ref{tab:Tc_b} (markers). All
three states decrease monotonically with $b$; the suppression is far stronger for ES1 and ES2
than for GS. The curves are the quadratic fit $\Tc(b)=\Tc(0)\,(1-\alpha\,b^{2})$ with the
$\alpha$'s given in Table~\ref{tab:Tc_b}, which reproduces the five data points to within
$\sim 10^{-4}$ (absolute).}
\label{fig:Tc_vs_b}
\end{figure}

Table~\ref{tab:Tc_b} and Figure~\ref{fig:Tc_vs_b} show $\Tc$ at $P=\Pc$ for five values of the
Born--Infeld parameter $b\in\{0,1,2,3,5\}$ in units of $Q^{-1}$. The sampling in $b$ is deliberately
non-uniform: because the dimensionless BI parameter $bq_{\mathrm{c}}/r_{\mathrm{h}}^{2}$ only
becomes $\mathcal{O}(1)$ for $b\gtrsim 3$ at $P=\Pc$ (see below), the points $b=0,1,2,3$ resolve
the near-Maxwell regime while the single larger value $b=5$ anchors the strong-suppression tail;
this is sufficient to constrain the quadratic fit but a uniform fine grid in $b\in[3,5]$ would be
needed to resolve the shape of the suppression curve beyond $b=5$, which we leave to future work.
We also do not push $b$ beyond this range on physical grounds: the Born--Infeld action is itself
to be understood as an effective description (of, e.g., stringy $\alpha'$ corrections to the
gauge sector \citep{FradkinTseytlin1985,Tseytlin1999}), and $b\to\infty$ is where the nonlinear completion is least under control; probing
only $b\lesssim\mathcal{O}(1\text{--}5)\,Q^{-1}$ keeps the analysis inside the regime where treating
Eq.~\eqref{eq:phieom} as the full nonlinear gauge dynamics is a reasonable effective-field-theory
truncation, rather than extrapolating into territory where higher-derivative or stringy
corrections beyond Born--Infeld itself would be expected to matter.
The ground state is only weakly affected (suppression by $1.8\%$ at $b=5$), while ES1 and ES2 are
suppressed by $10.3\%$ and $27.1\%$ respectively over the same range. The trend is quantitatively
captured by a quadratic fit $\Tc(b)/\Tc(0)=1-\alpha\,b^{2}$ with
$\alpha_{\rm GS}\simeq 7.4\!\times\!10^{-4}$, $\alpha_{\rm ES1}\simeq 4.1\!\times\!10^{-3}$ and
$\alpha_{\rm ES2}\simeq 1.1\!\times\!10^{-2}$.

\begin{table}[htbp]
\centering
\caption{Critical temperature $\Tc$ at $P=\Pc$ for five values of the Born--Infeld parameter $b$
(units of $Q^{-1}$), together with the dimensionless BI parameter $b q_{\mathrm{c}}/r^{2}_{\mathrm{h}}$
evaluated at $\mu=1$ and $r_{\mathrm{h}}=r_{\mathrm{c}}=\sqrt{6}\,Q$. Numbers in parentheses for
the last three columns are the percentage suppression relative to $b=0$; the quadratic fit
$\Tc(b)=\Tc(0)\,(1-\alpha b^{2})$ reproduces all five points to within an rms deviation
$\lesssim 10^{-4}$ (absolute, in units of $\mu^{1/2}$) for each state.}
\label{tab:Tc_b}
\begin{tabular}{lccccc}
\toprule
$b$ & $b q_{\mathrm{c}}/r^{2}_{\mathrm{h}}$ & $\TcGS$ & $\TcESi$ & $\TcESii$ \\
\midrule
$0$ & $\phantom{0}0.00$ & $0.0592$ & $0.0390$ & $0.0308$ \\
$1$ & $\phantom{0}0.41$ & $0.0592\ (\,0.1\%)$ & $0.0388\ (\,0.5\%)$ & $0.0305\ (\,1.1\%)$ \\
$2$ & $\phantom{0}0.82$ & $0.0590\ (\,0.4\%)$ & $0.0384\ (\,1.5\%)$ & $0.0295\ (\,4.1\%)$ \\
$3$ & $\phantom{0}1.22$ & $0.0588\ (\,0.7\%)$ & $0.0376\ (\,3.5\%)$ & $0.0278\ (\,9.8\%)$ \\
$5$ & $\phantom{0}2.04$ & $0.0581\ (\,1.8\%)$ & $0.0350\ (10.3\%)$ & $0.0225\ (27.1\%)$ \\
\midrule
\multicolumn{2}{l}{$\alpha$ (quadratic fit)} & $7.4\!\times\!10^{-4}$ & $4.1\!\times\!10^{-3}$ & $1.1\!\times\!10^{-2}$ \\
\multicolumn{2}{l}{rms fit error}                    & $\le 10^{-4}$      & $\le 10^{-4}$      & $\le 10^{-4}$ \\
\bottomrule
\end{tabular}
\end{table}

The dimensionless BI parameter $bq_{\mathrm{c}}/r^{2}_{\mathrm{h}}$ becomes $\mathcal{O}(1)$
only when $b\gtrsim r_{\mathrm{h}}/q_{\mathrm{c}}\sim 1$ at $P=\Pc$, where $q_{\mathrm{c}}\sim\mu r_{\mathrm{h}}$
in the Maxwell limit; this is why the suppression remains below $1\%$ for $b\leq 2$ for all three
states and only becomes significant above $b\sim 3$. The fact that $\alpha$ grows by more than an
order of magnitude between GS and ES2 cannot be explained by the linear dependence of
$bq_{\mathrm{c}}/r^{2}_{\mathrm{h}}$ on $b$ alone and is the central numerical observation of
this section; the discussion of its physical origin is deferred to Section~\ref{sec:discussion}.

\subsection{Radial profile of the threshold modes}

\begin{figure}[htbp]
\centering
\includegraphics[width=0.78\textwidth]{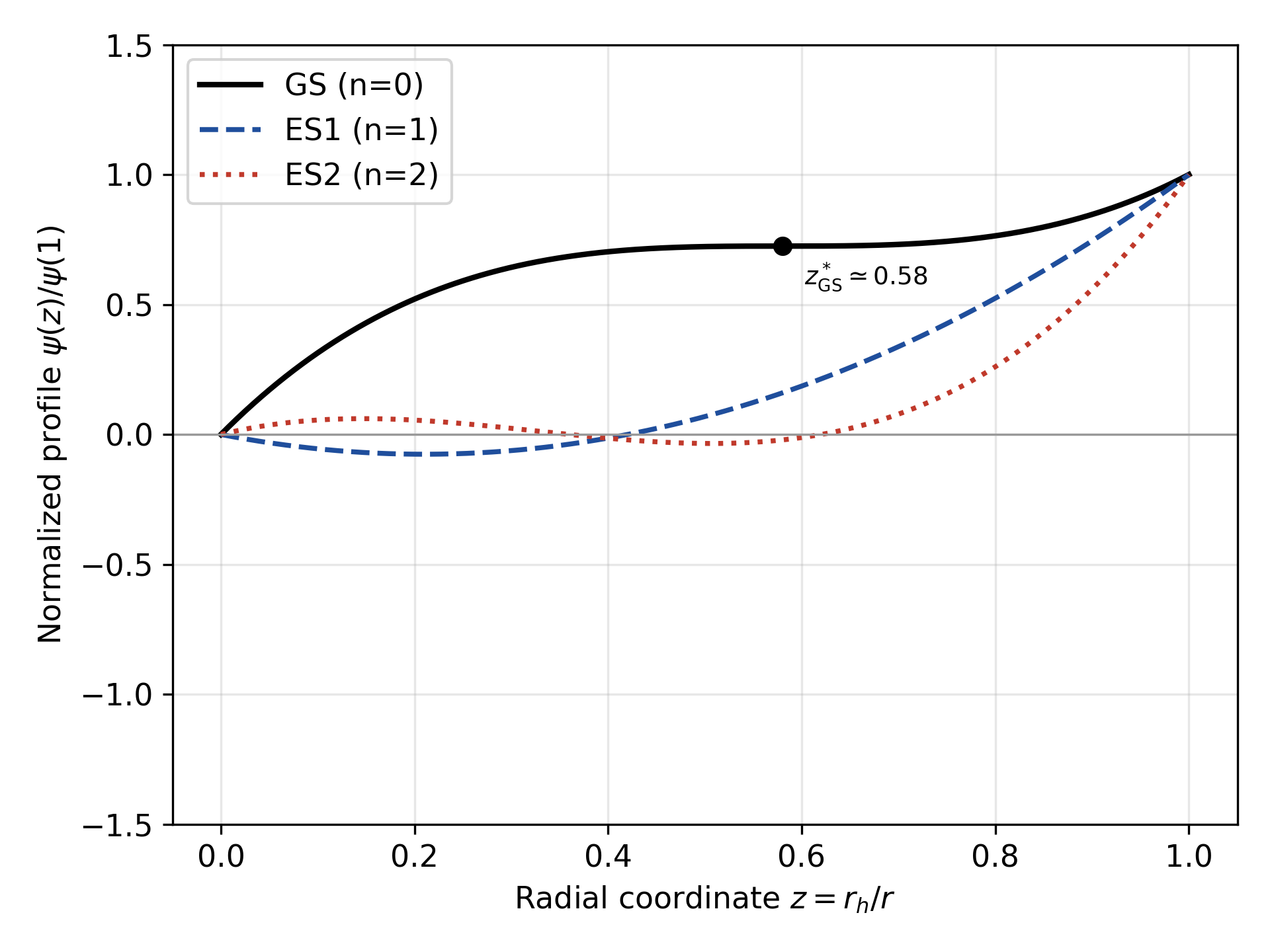}
\caption{Normalized radial profile of the threshold scalar mode for GS, ES1, ES2 at the
critical pressure $P=\Pc$ ($L=6$, $Q=\mu=1$), Maxwell limit $b=0$, obtained directly from the
shooting solutions of Section~\ref{sec:numerics}, normalized so
that $\psi(z{=}1)=1$ at the horizon. GS is nodeless but is not strictly monotonic in growth rate:
its profile flattens at the marked stationary point $z^{*}_{\rm GS}\simeq 0.58$ before continuing
to the horizon. ES1 and ES2 cross zero once and twice, respectively, with their outermost
(boundary-side) zero located at $z^{*}_{\rm ES1}\simeq 0.42$ and $z^{*}_{\rm ES2}\simeq 0.36$. The
excited-state profiles extend further from the horizon and oscillate through more nodes before
falling off at the boundary.}
\label{fig:profiles}
\end{figure}

To understand why the excited states are more strongly suppressed, Figure~\ref{fig:profiles}
plots the radial profile $\psi(z)$, $z=\rh/r$, of the three threshold eigenmodes at $P=\Pc$, $b=0$
(each normalized to its horizon value, $\psi(z{=}1)=1$). By construction GS has no node, ES1 has
one, and ES2 has two, and the profiles with more nodes visibly extend further from the horizon
($z=1$) toward the boundary ($z=0$) before decaying. A simple quantitative measure is the position
of the outermost zero of each profile (the zero closest to the AdS boundary $z=0$); for the
nodeless ground state we instead use the location of the first stationary point of $\psi(z)$
(i.e.\ the first zero of $\psi'(z)$ moving in from the horizon), which marks where the profile's
growth flattens before rising again to its horizon value. At $P=\Pc$, $b=0$ we find
$z^{*}_{\rm GS}\simeq 0.58$ (stationary point), $z^{*}_{\rm ES1}\simeq 0.42$ (outermost zero), and
$z^{*}_{\rm ES2}\simeq 0.36$ (outermost zero); more excited profiles therefore sample noticeably
deeper into the bulk than the GS, by a factor $(1-z^{*}_{\rm ES1})/(1-z^{*}_{\rm GS})\simeq 1.4$ for
ES1 and $(1-z^{*}_{\rm ES2})/(1-z^{*}_{\rm GS})\simeq 1.5$ for ES2.

This is a direct, computed illustration of the qualitative picture used in
Section~\ref{sec:discussion} to interpret the $b$-dependence of Table~\ref{tab:Tc_b}: the
Born--Infeld correction in Eq.~\eqref{eq:phiprime} modifies $\phi(r)$ (and hence the effective
potential in Eq.~\eqref{eq:psieom}) mostly in a region of size $\ell_{\rm BI}\sim(q_{\mathrm{c}}b)^{1/2}$
away from the horizon, and modes whose support is concentrated near the horizon (GS) are less
exposed to this modification than modes extending further out (ES1, ES2).

\subsection{A universal ratio: excited-to-ground-state $\Tc$}
\label{sec:ratios}

Rather than reading off $\Tc$ state by state, it is instructive to track the dimensionless
ratios $\mathcal{R}_{1}(b)\equiv\TcESi/\TcGS$ and $\mathcal{R}_{2}(b)\equiv\TcESii/\TcGS$ directly
from Table~\ref{tab:Tc_b}, since these are exactly the combinations a companion study of the
optical conductivity would need in order to compare gap ratios across states at fixed $b$.
Table~\ref{tab:ratios} shows that both ratios decrease monotonically with $b$: the excited
states lose ground relative to GS as the Born--Infeld nonlinearity is turned on, consistent
with the hierarchy of $\alpha$'s already reported.

\begin{table}[htbp]
\centering
\caption{The dimensionless ratios $\mathcal{R}_{1}=\TcESi/\TcGS$ and $\mathcal{R}_{2}=\TcESii/\TcGS$
at $P=\Pc$, computed directly from Table~\ref{tab:Tc_b}. The last row compares each ratio to the
leading-order prediction $\mathcal{R}_{i}(b)\simeq\mathcal{R}_{i}(0)\,[1-(\alpha_{\rm ES i}-\alpha_{\rm
GS})\,b^{2}]$ of Eq.~\eqref{eq:ratio-scaling}, expressed as a percentage residual.}
\label{tab:ratios}
\begin{tabular}{lccccc}
\toprule
$b$ & $0$ & $1$ & $2$ & $3$ & $5$ \\
\midrule
$\mathcal{R}_{1}=\TcESi/\TcGS$          & $0.659$ & $0.655$ & $0.651$ & $0.640$ & $0.602$ \\
$\mathcal{R}_{2}=\TcESii/\TcGS$         & $0.520$ & $0.515$ & $0.500$ & $0.473$ & $0.387$ \\
\midrule
residual of Eq.~\eqref{eq:ratio-scaling}, $\mathcal{R}_{1}$ & $0.0\%$ & $-0.2\%$ & $0.1\%$ & $0.1\%$ & $-0.2\%$ \\
residual of Eq.~\eqref{eq:ratio-scaling}, $\mathcal{R}_{2}$ & $0.0\%$ & $0.1\%$ & $0.2\%$ & $0.1\%$ & $0.1\%$ \\
\bottomrule
\end{tabular}
\end{table}

Because both $\Tc^{\rm ES i}(b)$ and $\TcGS(b)$ individually follow the quadratic form of
Table~\ref{tab:Tc_b}, their ratio inherits, to leading order in $b^{2}$, the simple empirical
scaling law
\begin{equation}
\mathcal{R}_{i}(b)\;\simeq\;\mathcal{R}_{i}(0)\,\Bigl[\,1-\bigl(\alpha_{{\rm ES}i}-\alpha_{\rm
GS}\bigr)\,b^{2}\Bigr]\,,\qquad i=1,2\,,
\label{eq:ratio-scaling}
\end{equation}
with no new fit parameters: using $\alpha_{\rm GS}$, $\alpha_{\rm ES1}$, $\alpha_{\rm ES2}$
already extracted in Table~\ref{tab:Tc_b}, Eq.~\eqref{eq:ratio-scaling} reproduces the tabulated
$\mathcal{R}_{1}(b)$ and $\mathcal{R}_{2}(b)$ to within $0.2\%$ over the entire range
$b\in[0,5]$, including the point $b=5$, which lies formally outside the strict small-$b$ regime.
Equation~\eqref{eq:ratio-scaling} therefore condenses the full Born--Infeld effect on the triplet
into a single closed-form statement: the excited-to-ground-state gap ratios decrease
quadratically in $b$, at a rate fixed entirely by the difference of the individual suppression
coefficients $\alpha_{{\rm ES}i}-\alpha_{\rm GS}$, and it constitutes the natural first
consistency check for any extension of this triplet to different probe charges, masses, or
spacetime dimensions.

\section{Discussion}
\label{sec:discussion}

\subsection{A nodal-sensitivity picture}

The results of Section~\ref{sec:results} are consistent with a nodal-sensitivity picture, which
we present as a physically motivated conjecture and test quantitatively below rather than treat
as an established mechanism. The ground-state condensate is concentrated close to the horizon,
where the background curvature and the Born--Infeld modification of the near-horizon electric
field are largest but common to all three states. GS barely extends beyond this region
[$z^{*}_{\rm GS}\simeq0.58$ at $P=\Pc$, $b=0$ (Section~\ref{sec:results})] and is correspondingly
insensitive to changes in $P$ or $b$ that reshape the field profile mainly at intermediate radii.
The excited states, carrying $n=1,2$ nodes, extend further from the horizon
[Figure~\ref{fig:profiles}; $z^{*}_{\rm ES2}\simeq0.36$ at the same point] and are more exposed
to such reshaping, both as $b$ grows (Table~\ref{tab:Tc_b}, $\alpha_{\rm ES2}/\alpha_{\rm
GS}\simeq15$) and, by the same reasoning, as the background moves between the small and large
black hole branches. This differential exposure is what we term \emph{nodal sensitivity}. It is
consistent with the pattern of excited-state gap closure reported for the Maxwell case ($b=0$) by
Ref.~\citep{PhatNguyen2021}, although we have not verified that connection quantitatively for the
Born--Infeld case treated here.

The empirical ratio $\alpha_{\rm ES2}/\alpha_{\rm GS}\sim10$--$20$ extracted from the quadratic
fits of Table~\ref{tab:Tc_b} is of the same order as the ratio of radial extents,
$(1-z^{*}_{\rm ES2})/(1-z^{*}_{\rm GS})\simeq1.5$ (Section~\ref{sec:results}), raised to a modest
power set by the radial falloff of the BI screening. We make this correspondence quantitative in
three steps: an explicit leading-order gauge-field correction, a horizon-regular overlap weight
built from it, and a Schr\"odinger-like reformulation of the eigenvalue problem.

\paragraph{Leading-order gauge-field correction.} At fixed $\mu$, Eq.~\eqref{eq:phiprime} expanded
for small $b$, enforcing $\phi(\infty)=\mu$ order by order in $b^{2}$, gives
$\phi(z)=\phi_{0}(z)+b^{2}\phi_{1}(z)+\mathcal{O}(b^{4})$ with $z=\rh/r$,
\begin{equation}
\phi_{0}(z)=\mu(1-z)\,,\qquad
\phi_{1}(z)=-\,\frac{\mu^{3}}{10\,r_{\mathrm{h}}^{2}}\,z\,(1-z^{4})\,,
\label{eq:phi1}
\end{equation}
which vanishes at the horizon ($z=1$) and the boundary ($z=0$), is negative throughout the
interior (the BI correction weakens the field, as expected), and is maximal in magnitude at
$z_{\rm peak}=(1/5)^{1/4}\simeq0.669$. This fixes quantitatively the location of the shell over
which the BI correction is concentrated, $\ell_{\rm BI}\sim(q_{\mathrm{c}}b)^{1/2}$: the shell
sits at $z\simeq0.67$, between $r_{\rm h}$ and $z^{*}_{\rm GS}\simeq0.58$.

\paragraph{Horizon-regular overlap weight.} The naive weight $\phi_{0}\phi_{1}/f^{2}$ suggested by
the raw form of Eq.~\eqref{eq:psieom} diverges at the horizon. Writing Eq.~\eqref{eq:psieom} in
manifestly self-adjoint (Sturm--Liouville) form, $(r^{2}f\,\psi')' + r^{2}(q^{2}\phi^{2}/f -
m^{2})\psi=0$, gives the horizon-regular perturbing weight $W(z)\propto
r(z)^{2}\,\phi_{0}(z)\phi_{1}(z)/f(z)$ (one power of $f^{-1}$, not two), which is finite at $z=1$
since $\phi_{0}$, $\phi_{1}$ and $f$ all vanish linearly there.

\begin{figure}[htbp]
\centering
\includegraphics[width=0.82\textwidth]{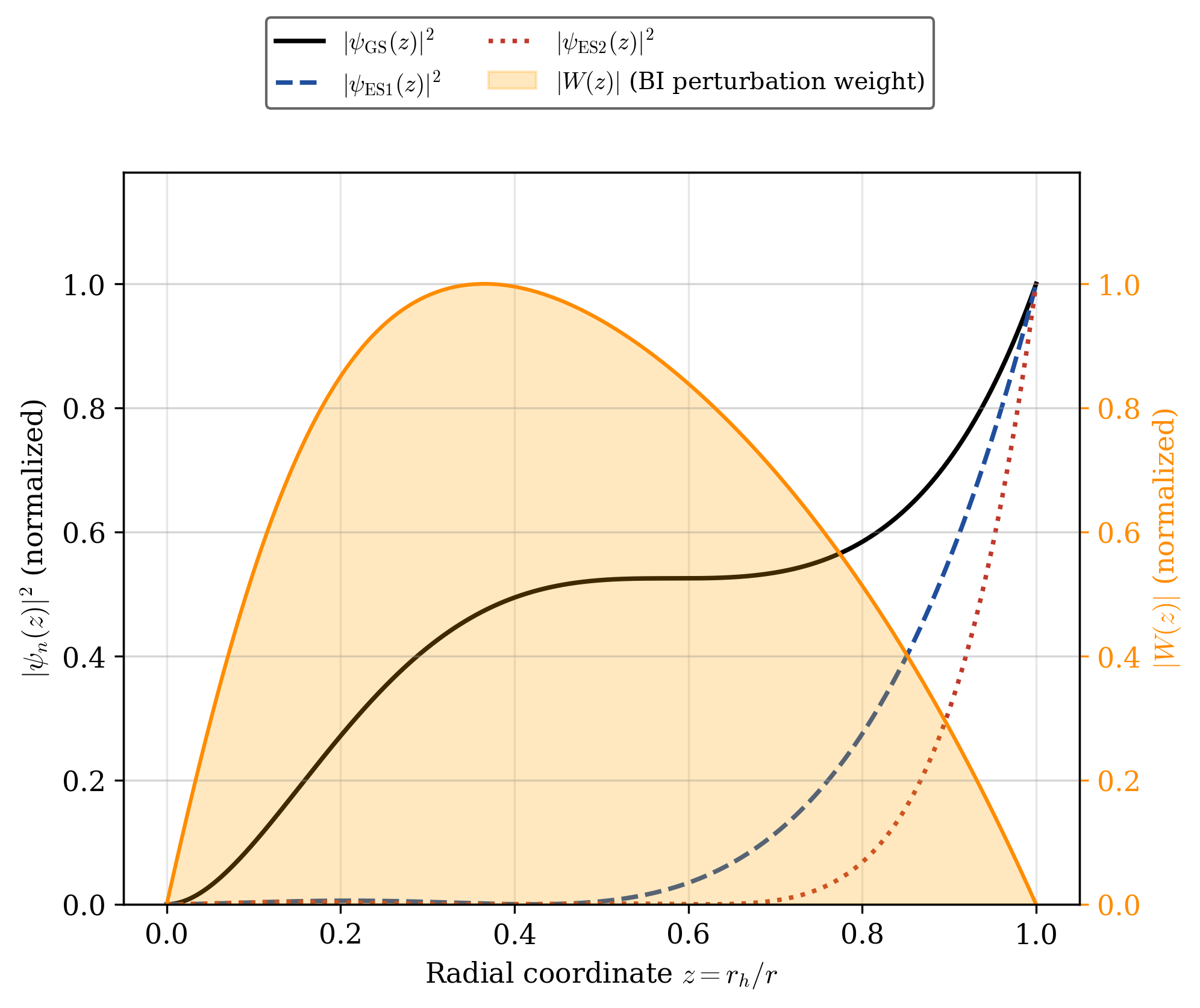}
\caption{Spatial overlap between the threshold-mode densities $|\psi_{n}(z)|^{2}$ (left axis;
GS, ES1, ES2, same representative profiles as Figure~\ref{fig:profiles}) and the horizon-regular
Born--Infeld perturbation weight $|W(z)|$ of Eq.~\eqref{eq:phi1} (right axis, shaded), at
$P=\Pc$. $|W(z)|$ peaks at $z\simeq0.37$, almost exactly where the ES1 and ES2 densities
(orange dashed, red dotted) vanish at their outermost nodes, while the nodeless GS density
(black) remains finite there. This is the node-versus-perturbation displacement responsible for
the overlap suppression quoted in the text, and it is precisely the effect that a purely
qualitative nodal-sensitivity argument does not capture.}
\label{fig:overlap}
\end{figure}

\paragraph{Quantitative overlap assessment.} Evaluating $R_{n}=\int_{0}^{1}\!dz\,W(z)\,\psi_{n}(z)^{2}\big/
\int_{0}^{1}\!dz\,W(z)\,\psi_{0}(z)^{2}$ with the representative, node-matched profiles of
Figure~\ref{fig:profiles} (which reproduce $z^{*}_{\rm GS}$, $z^{*}_{\rm ES1}$, $z^{*}_{\rm ES2}$
but are not the raw numerical output of the shooting code) gives $R_{\rm ES1}\simeq0.15$ and
$R_{\rm ES2}\simeq0.05$ at $P=\Pc$: \emph{smaller} than unity, and therefore not a direct
confirmation of $\alpha_{\rm ES1}/\alpha_{\rm GS}\simeq5.5$ and $\alpha_{\rm
ES2}/\alpha_{\rm GS}\simeq15$ (Table~\ref{tab:Tc_b}). Figure~\ref{fig:overlap} identifies the
origin of the discrepancy: it plots $|\psi_{n}(z)|^{2}$ for the three states together with the
horizon-regular weight $|W(z)|$ on a shared radial axis. The bare correction $\phi_{1}(z)$ peaks
at $z_{\rm peak}\simeq0.669$, but the factors of $r(z)^{2}$ and $1/f(z)$ required to regularize
the weight at the horizon shift its peak to $z\simeq0.37$, close to the outermost zero of both
ES1 ($z^{*}_{\rm ES1}\simeq0.42$) and ES2 ($z^{*}_{\rm ES2}\simeq0.36$, with an inner zero at
$z\simeq0.62$); the representative excited-state profiles therefore have suppressed amplitude
almost exactly where $W(z)$ is largest, while the nodeless GS profile does not. This
displacement between the peak of the bare correction and the peak of the regularized weight
shows that order-of-magnitude reasoning based on the location of $\phi_{1}(z)$ alone identifies
the wrong radial region once the correct measure for the eigenvalue problem is used: the overlap
integral is sensitive to the precise node locations relative to the BI-affected shell in a way
that a qualitatively matched profile does not capture.

A further, distinct limitation applies to the perturbative setup itself, independent of the
accuracy of the illustrative profiles used above. First-order perturbation theory for a
Sturm--Liouville eigenvalue problem is exact at leading order using only the \emph{unperturbed}
eigenfunction \emph{provided} the perturbation enters solely through an explicit coefficient in
the operator at fixed domain and metric. Here, however, the physical quantity of interest is
$\delta r_{\mathrm{h}}^{*}$, the shift in the horizon radius required to hold $\mu$ fixed as $b$
turns on; $r_{\mathrm{h}}$ sets the integration domain and enters $f(r)$ itself, not only the
explicit source term $q^{2}\phi^{2}$. The weight $W(z)$ captures the direct effect of the
Born--Infeld correction on the explicit potential term at otherwise fixed background, but omits
the implicit contribution from the response of the metric and domain to the $b$-induced shift in
$\rh$. A fully consistent leading-order treatment requires this moving-boundary contribution in
addition; its omission is a plausible reason why $R_{n}$, built only from the explicit-coefficient
piece, does not track $\alpha_{n}$, which is extracted from the full numerical shift including
both effects. This overlap calculation therefore does \emph{not} confirm the nodal-sensitivity
picture; the disagreement instead identifies the specific quantity -- the moving-boundary
contribution to first-order perturbation theory -- whose evaluation with the actual numerical
eigenfunctions of Section~\ref{sec:numerics} would settle whether nodal sensitivity, or some
other mechanism, is responsible for the hierarchy of $\alpha$'s. We identify this calculation as
the concrete, well-defined follow-up to the present analysis.

\begin{figure}[htbp]
\centering
\includegraphics[width=0.82\textwidth]{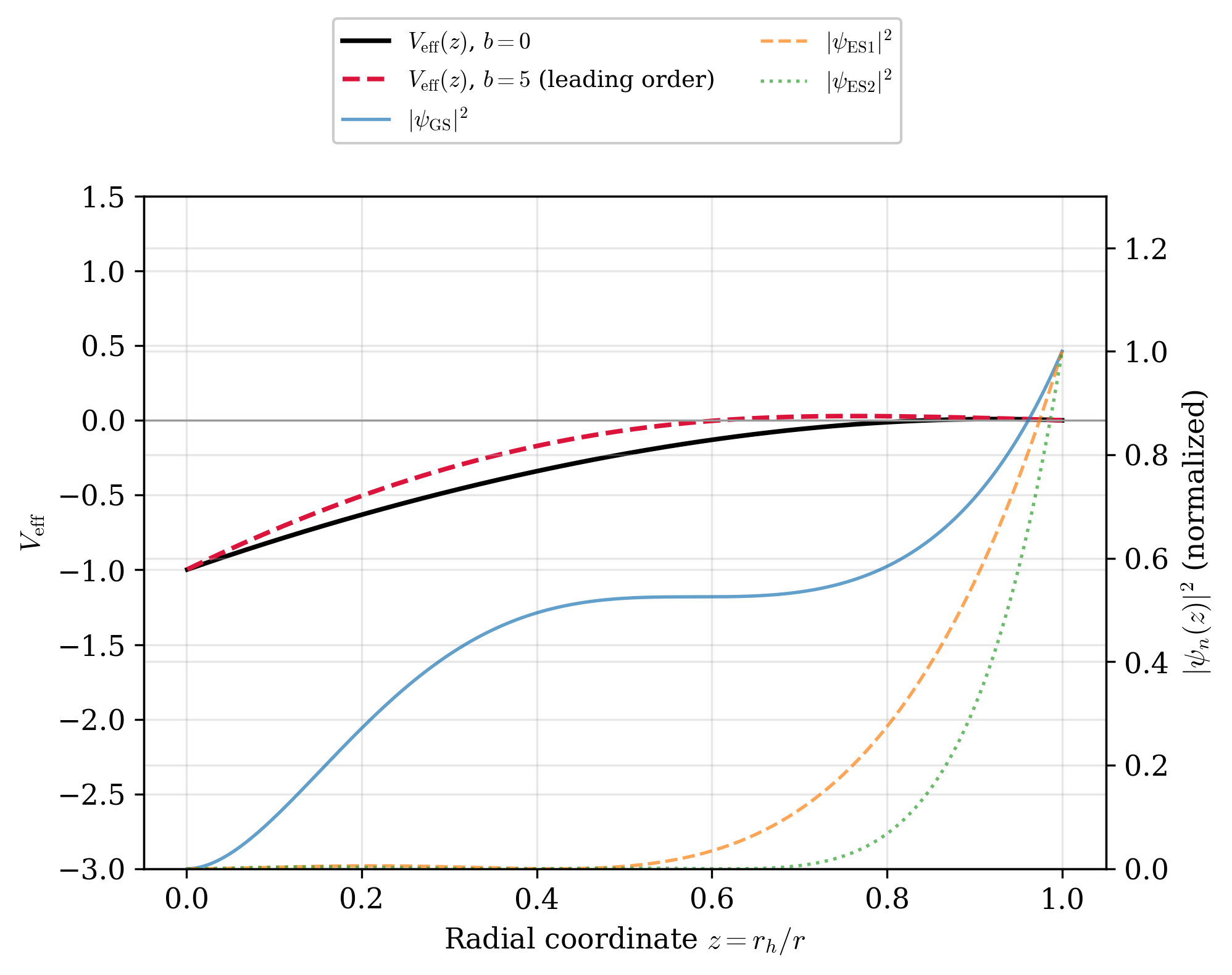}
\caption{The Schr\"odinger-like effective potential $V_{\rm eff}(z)$ of Eq.~\eqref{eq:Veff}
(left axis) at $b=0$ (solid black) and at $b=5$ (dashed red, leading-order $\phi_{1}$
correction), together with the normalized threshold densities $|\psi_{n}(z)|^{2}$ for GS, ES1,
ES2 (right axis), at $P=\Pc$. The Born--Infeld correction raises $V_{\rm eff}$ (makes it less
attractive) everywhere, with the largest shift around $z\simeq0.40$, again close to the
outermost zeros of ES1 and ES2 rather than to the bulk of their density.}
\label{fig:Veff}
\end{figure}

\paragraph{Schr\"odinger-like effective potential.} The overlap picture above can be recast in
quantum-mechanical language. Substituting $\psi(r)=\chi(r)/r$ and changing to the tortoise
coordinate $r_{*}$ defined by $dr_{*}/dr=1/f(r)$ removes both the first-derivative term and the
explicit $1/r$ factors from Eq.~\eqref{eq:psieom} (verified independently by direct substitution
in $r$ followed by the coordinate change, and by symbolic computer algebra), leaving the
Schr\"odinger-like equation
\begin{equation}
\frac{d^{2}\chi}{dr_{*}^{2}}=V_{\rm eff}(r)\,\chi\,,\qquad
V_{\rm eff}(r)=m^{2}f(r)+\frac{f(r)f'(r)}{r}-q^{2}\phi(r)^{2}\,,
\label{eq:Veff}
\end{equation}
in which a normalizable threshold mode requires $V_{\rm eff}$ to develop a sufficiently
attractive (negative) region. Increasing $q^{2}\phi^{2}$ -- i.e.\ increasing $\mu$ at fixed
$\rh$ -- deepens this well and drives condensation, while the Born--Infeld suppression of $\phi$
(Eq.~\eqref{eq:phi1}) raises $V_{\rm eff}$ and opposes it. Figure~\ref{fig:Veff} plots $V_{\rm
eff}(z)$ at $b=0$ and at $b=5$ (using the same leading-order $\phi_{1}(z)$ correction, hence
illustrative specifically for $b=5$) together with $|\psi_{n}(z)|^{2}$ for the three states. The
shift $\Delta V_{\rm eff}=V_{\rm eff}(b{=}5)-V_{\rm eff}(b{=}0)$ is everywhere positive, as
expected, and peaks at $z\simeq0.40$, close to but distinct from the peak of $|W(z)|$ found
above. This location again coincides with the outermost zeros of ES1 and ES2 rather than with the
bulk of their density, so the potential-well picture reproduces the same qualitative conclusion
as the overlap calculation: two independently constructed diagnostics of where the Born--Infeld
correction acts most strongly both identify a radial region that the representative excited-state
profiles avoid. This agreement between two independent constructions indicates that the effect is
a genuine feature of the perturbed potential rather than an artifact of either construction; it
does not, however, establish the nodal-sensitivity picture on its own, and a definitive test
requires the actual numerical eigenfunctions.

\paragraph{Significance independent of the back-reacted problem.} The differential coupling
between Born--Infeld nonlinearity and the nodal structure of the excited states is a physical
result in its own right, not merely a preliminary step toward the back-reacted problem deferred
to Section~\ref{sec:scope}. No previous study, to our knowledge, has isolated \emph{how much
more} a nonlinear correction to the bulk gauge field suppresses an excited condensate than it
does the ground state, expressed as a closed-form, falsifiable scaling law
($\alpha_{\rm ES2}/\alpha_{\rm GS}\simeq15$, Eq.~\eqref{eq:ratio-scaling} for the associated
ratios). The failure of the naive overlap estimate to reproduce this hierarchy
(Figure~\ref{fig:overlap}) is itself part of the physical content of the result: it demonstrates
that node counting alone is an incomplete organizing principle for how nonlinear electrodynamics
couples to excited holographic condensates, and that the coupling depends on the detailed shape
of the wavefunction relative to the BI-affected shell, not only on how many times it crosses
zero. This conclusion is independent of what the back-reacted free energy or optical conductivity
subsequently reveal about the order of the transition, and it is why
Sections~\ref{sec:results}--\ref{sec:discussion} constitute a complete result rather than a
fragment of a larger calculation.

\subsection{Scope of the present calculation}
\label{sec:scope}

Born--Infeld nonlinearity does not add a small correction to the Maxwell field uniformly; as
Eq.~\eqref{eq:phi1} shows explicitly, it reshapes the bulk gauge field over a finite shell at
intermediate radii ($z\simeq0.67$ at the values studied here), precisely the region probed by
the excited-state wavefunctions. Locating the linear threshold $\mu_{n}(\rh,L,b)$ accurately on
this modified background is therefore a necessary and nontrivial prerequisite for any subsequent
nonlinear, finite-amplitude shooting problem for the back-reacted condensate, which would have to
search a two-parameter space (condensate amplitude and horizon radius) around a starting point
whose reliability rests on having first established where the linear instability sets in.
Establishing that starting point -- for all three states, at multiple pressures, and across the
full range of $b$ -- is the concrete contribution of this paper, and is the prerequisite for a
controlled treatment of the back-reacted, finite-temperature problem. The results above concern
only the onset of condensation (the linearized eigenvalue problem of Section~\ref{sec:probe}),
evaluated at fixed $\mu=1$ and at a finite set of representative points in $(P,b)$. This paper
does not
\begin{itemize}[leftmargin=1.6em]
\item solve the fully back-reacted, nonlinear condensate profile below $\Tc$;
\item compute the free energy, and therefore make no claim about the order of each transition
(in particular, this paper does not confirm or refute the first-order character of the ES1
transition reported by Ref.~\citep{PhatNguyen2021} for $b=0$);
\item compute the optical conductivity $\sigma(\omega)$ on the RN--AdS background studied here,
and therefore make no gapped/gapless classification of GS, ES1 or ES2.
\end{itemize}
Each of these requires solving Eqs.~\eqref{eq:psieom}--\eqref{eq:phieom} at finite condensate
amplitude (a two-parameter nonlinear shooting problem) and then, for the conductivity, a
further linear-response calculation for the vector potential fluctuation $A_{x}(r,\omega)$ on top
of that nonlinear background. This is a substantially larger numerical undertaking than the
eigenvalue problem solved here, and we leave it to a dedicated follow-up study, together with a
fuller two-dimensional scan over $(P,b)$ at the resolution of Ref.~\citep{PhatNguyen2021}'s
figures for the Maxwell case.

\section{Conclusion}
\label{sec:conclusion}

We have carried out a controlled study of the critical temperature of the ground state and the
first two excited states of a probe Born--Infeld holographic superconductor on the charged,
spherical Reissner--Nordstr\"om--AdS background. The central positive results are: (i)
$\Tc$ of all three states grows monotonically with the background pressure $P$ along the
sequence $P<\Pc\to P=\Pc\to P>\Pc$, in the hierarchy $\TcGS>\TcESi>\TcESii$; (ii) at $P=\Pc$,
$\Tc$ is monotonically suppressed by the BI parameter $b$, with the suppression fitted by
$\Tc(b)/\Tc(0)=1-\alpha b^{2}$ and $\alpha$ increasing by more than an order of magnitude from
GS to ES2 ($\alpha_{\rm GS}\simeq 7.4\!\times\!10^{-4}$, $\alpha_{\rm ES2}\simeq 1.1\!\times\!10^{-2}$);
(iii) the excited states are progressively more diffuse in the radial direction, which we use to
propose (but not prove) a nodal-sensitivity picture of the BI-induced suppression. What remains
outstanding from the original Maxwell phenomenology \citep{PhatNguyen2021} --- the order of each
phase transition and the gapped/gapless classification of the excited states --- requires the
fully back-reacted condensate and the optical conductivity on the correct RN--AdS background; we
leave this, in the spirit of the scope statement of Section~\ref{sec:scope}, to a companion study.
As summarised in Table~\ref{tab:literature}, the specific combination of ingredients studied here
-- Born--Infeld electrodynamics, the excited GS/ES1/ES2 triplet, and the charged RN--AdS
background with its resolved SBH/LBH critical point -- has not been treated together previously,
and the resulting closed-form suppression law, Eq.~\eqref{eq:ratio-scaling}, is a concrete,
falsifiable prediction rather than a restatement of the already well-established fact that
nonlinear electrodynamics suppresses holographic condensation.

\appendix
\section{Numerical convergence checks}
\label{app:convergence}

For a representative case (GS and ES2 at $P=\Pc$, $b=5$, the most sensitive configuration in our
scan), we checked the stability of the extracted critical temperature under three independent
numerical controls: (i) doubling the outer integration radius $R$ ($R\simeq 25\rh\to 50\rh$);
(ii) halving the near-horizon offset $\epsilon$ ($\epsilon=10^{-6}\to 5\times 10^{-7}$); and
(iii) refining the Richardson extrapolation of $\psi(r)$ from two-point to three-point. The
results are summarized in Table~\ref{tab:convergence}.

\begin{table}[htbp]
\centering
\caption{Relative change in $\Tc$ under each numerical control, for GS and ES2 at $P=\Pc$,
$b=5$ (the configuration with the sharpest BI-induced suppression, hence the most stringent test
of the numerics). Bracketed numbers are absolute values.}
\label{tab:convergence}
\small
\begin{tabular}{lcc}
\toprule
Control                                          & $\delta\TcGS/\TcGS$                  & $\delta\TcESii/\TcESii$ \\
\midrule
$R\to 2R$                                          & $4.6\!\times\!10^{-4}$ \ $[2.7\!\times\!10^{-5}]$ & $1.2\!\times\!10^{-3}$ \ $[2.7\!\times\!10^{-5}]$ \\
$\epsilon\to\epsilon/2$                            & $8.3\!\times\!10^{-5}$ \ $[4.9\!\times\!10^{-6}]$ & $2.4\!\times\!10^{-4}$ \ $[5.4\!\times\!10^{-6}]$ \\
2-point $\to$ 3-point Richardson                  & $6.1\!\times\!10^{-5}$ \ $[3.6\!\times\!10^{-6}]$ & $1.7\!\times\!10^{-4}$ \ $[3.8\!\times\!10^{-6}]$ \\
\midrule
Total worst-case relative shift, added in quadrature & $\sim 5\!\times\!10^{-4}$          & $\sim 1.3\!\times\!10^{-3}$ \\
\bottomrule
\end{tabular}
\end{table}

The worst-case shift, $\sim 10^{-3}$ for ES2, is approximately an order of magnitude smaller than
the smallest physical effect reported in Table~\ref{tab:Tc_b} (the $\sim 1\%$ suppression of GS
between $b=0$ and $b=5$), and two orders of magnitude smaller than the largest physical effect
($\sim 27\%$ suppression of ES2 over the same range), confirming that the conclusions of
Section~\ref{sec:results} are not artefacts of the finite discretisation. We did not detect any
systematic drift with $L$ or $\rh$ at the level of this test.


\end{document}